\documentclass[onecollarge,natbib]{svjour2}
\bibpunct{[}{]}{;}{n}{}{,} 
\smartqed  
\usepackage{graphicx}
\usepackage{color}
\usepackage{amssymb,amsmath}

\newcommand{\aho}{a_\text{ho}}
\newcommand{\ve}{\mathbf}

\DeclareMathOperator{\sgn}{sgn}

\journalname{Few-Body Systems}
\begin{document}

\title{Efimov Trimers in a Harmonic Potential\thanks{Special issue devoted to Efimov physics}}

\author{Jacobus Portegies \and Servaas Kokkelmans}

\institute{Eindhoven University of Technology, P.O. Box 513, 5600 MB Eindhoven, The Netherlands\\
\email{jim@cims.nyu.edu, s.kokkelmans@tue.nl} \\
\emph{Present address of Jacobus Portegies:} Courant Institute of Mathematical Sciences, 251 Mercer Street, New York, NY 10012-1185, United States
}


\date{Received: date / Accepted: date}

\maketitle

\begin{abstract}
We study the Efimov effect in a harmonic oscillator in the hyperspherical formulation, and show how a reduced model allows for a description that is a generalization of the Efimov effect in free space and leads to results that are easily interpreted. Efimov physics may be observed by varying the value of the scattering length, since in the regime where the trimers have a mixed harmonic oscillator and Efimov character, the inelastic properties of these states are still manageable. The model also allows for the study of non-universal Efimov trimers by including the effective range scattering parameter. While we find that in a certain regime the effective range parameter can take over the role of the three-body parameter, interestingly, we obtain a numerical relationship between these two parameters different from what was found in other models.

\end{abstract}

\section{Introduction}

In 1970, Efimov described a phenomenon occurring in a three-particle system of identical bosons, in which there is an accumulation of bound states near the point of zero energy and diverging scattering length of the interparticle interaction \cite{Efimov}. Efimov was inspired by the physics of the nucleus of the tritium atom. However, the experimental signature for the existence of such Efimov states was not observed in nuclear physics, but in a gas of ultracold cesium atoms~\cite{Kraemer06} by studying three-body recombination. The reason that properties predicted for nuclear physics few-body systems can be derived from ultracold atoms lies in the fact that the two systems are connected through universal relations, that are irrespective of the specific nature of the interparticle interactions.

\begin{figure}
\begin{center}
\includegraphics[width=8cm]{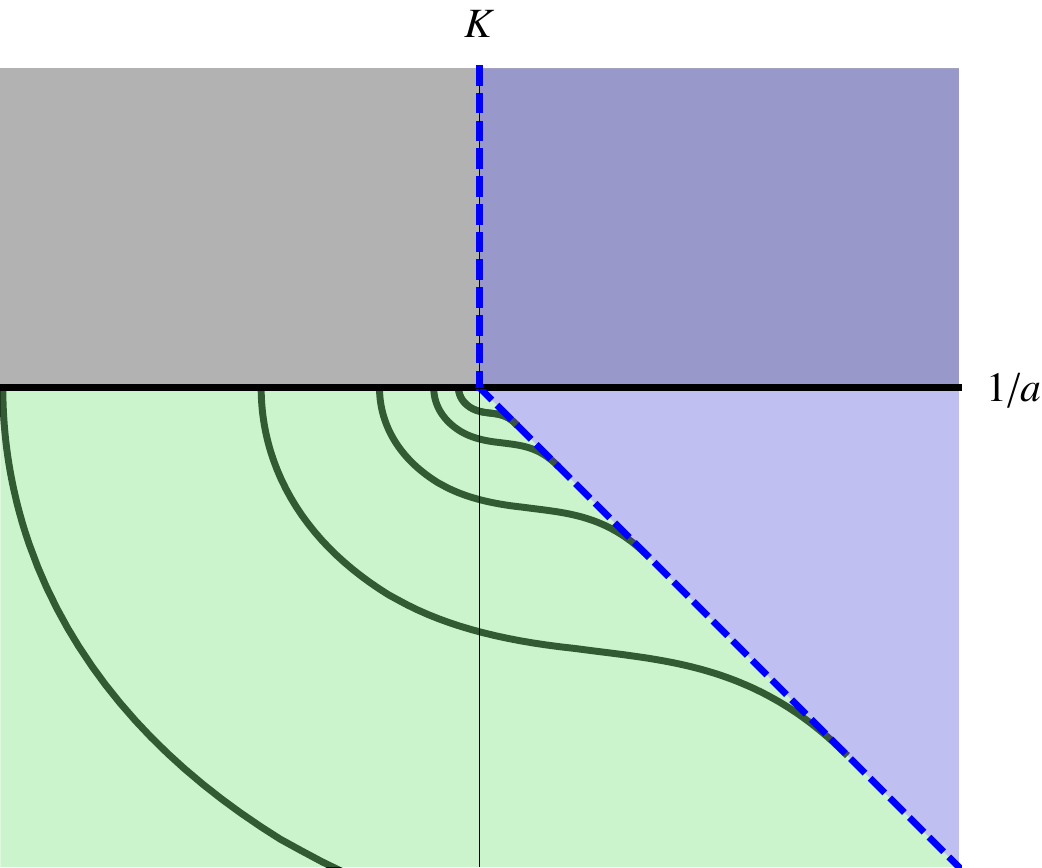}
\caption{A schematic description of the Efimov effect: shown is the wavenumber $K\sim \sqrt E$, indicating dimer and Efimov trimer states as a function of the inverse scattering length $1/a$. The horizontal line $K=0$ indicates the threshold to the three-body scattering continuum (gray). Simultaneously, the blue region marks the atom-dimer continuum for $a>0$, and the green zone indicates the trimer region. The dimer binding energy scales as $-1/a$. Only a few Efimov states are shown. The figure is not to scale: for clarity reasons we plotted instead of $K$ the quantity $H^{1/4} \sin \xi$, and instead of $1/a$ the quantity $H^{1/4} \cos \xi$ (cf.~Eq.~(\ref{eq:polarcord})).}\label{fi:efimovschematic}
\end{center}
\end{figure}

Since the first observations in cesium, the Efimov effect has received a lot of attention~\cite{Ferlaino10,greene10}. Several groups now have observed Efimov physics by studying three-body recombination in a variety of atomic systems~\cite{Knoop09,Zaccanti09,Gross09,Pollack09,Ottenstein08,Lompe10a,Huckans09,Williams09,Nakajima10,Barontini09}. Evidence for the existence of trimers appears as some distinct features (resonances) in the recombination rate coefficient. In Fig. \ref{fi:efimovschematic} the energies of the Efimov trimers are shown, as a function of the inverse two-body scattering length $1/a$.

A crucial advantage of ultracold atoms with respect to nuclear systems is that the interparticle interactions can be easily manipulated using Feshbach resonances~\cite{Feshbach58,Tiesinga92}, by varying the magnetic field. The two-body scattering length, which is obtained from the scattering phase shift $\delta(k)$ as $a=-\lim_{k \rightarrow 0} \tan \delta(k)/k$, diverges on resonance to infinity, and is typically described by the dispersive formula

\begin{equation}
 a=a_{bg} \left(1-\frac{\Delta B}{B-B_0} \right),
\end{equation}
with $a_{bg}$ the background value of the scattering length, $B$ the magnetic field, $B_0$ the resonance position, and $\Delta B$ the width of the resonance. This possibility to change the scattering length allowed for the observation of distinct Efimov features in the three-body recombination rate, connected to two different Efimov states in a row \cite{Zaccanti09,Pollack09}, confirming the scaling law~\cite{Efimov,Braaten&Hammer06} given by

\begin{equation}
 E_T^{n+1}/E_T^n = e^{-2\pi/s_0} \approx 1/515, 
\label{efimovscaling}
\end{equation}
with $s_0=1.00624$ a constant. Moreover, recent developments extended the universal few-body physics to the domain of four-body states which was theoretically predicted~\cite{Stecher09} and experimentally
verified~\cite{Pollack09,Ferlaino09}. 

In most of the experiments, Efimov physics has been observed only indirectly via three-body recombination~\cite{Braaten&Hammer06}. Due to the short lifetime of these states, it is difficult to probe the Efimov states directly. These weakly-bound trimers are not only unstable in collisions with other particles, but they are also intrinsically unstable by itself, as trimers may decay to a deeper lying dimer state and an unbound atom.
However, the group of Jochim in Heidelberg recently succeeded in associating Efimov trimers directly via a new and promising approach using radio-frequency fields~\cite{Lompe10b}.

Optical lattices might provide more experimental control, and potentially give rise to an enhanced lifetime of Efimov trimers. Efimov physics can be studied in this way by loading three atoms in each lattice site. When the lattice is deep enough, single sites act as harmonic oscillator potentials where individual trimers can be trapped. A possible advantage is then that the interparticle separation of the trimers can be increased to the scale of the harmonic oscillator length, while their properties are still determined by Efimov physics. In this way the overlap of these states with deeply bound dimers is reduced, thereby suppressing the process of three-body recombination. The influence of other limiting processes, where four or more particles involved, will be reduced as well. 

Jonsell, Heiselberg and Pethick were the first to introduce an harmonic oscillator term in a three-body system~\cite{Jonsell02}, by adapting the approach of Ref.~\cite{Fedorov93}. For diverging scattering length, they found exact expressions for the bound states of three identical bosons, while they obtained numerical results for finite scattering length. Later, Werner and Castin \cite{Werner06} obtained additional analytical results for diverging scattering length, and complimentary first-order corrections for large scattering length. Stoll and K\"{o}hler \cite{Stoll05} performed numerical simulations with a tunable separable potential to find the bound states of three particles in a trap. Th\o{}gersen {\it et al.}~found numerical solutions of three particles in a harmonic trap~\cite{Thogersen08}, mainly to study the non-universal corrections to the Efimov effect that come from the effective range. More recently, Liu, Hu and Drummond derived an analytical description of three attractively interacting fermions in a harmonic trap~\cite{Liu10}.

A convenient description of the Efimov effect in free space is presented in Ref.~\cite{Braaten&Hammer06}, in which polar coordinates $H$ and $\xi$ are introduced, which are related to the wave number or energy parameter $K$ and the scattering length $a$ via
\begin{subequations}
\label{eq:polarcord}
\begin{eqnarray}
K &= H \sin \xi,\\
\frac{1}{a} &= H \cos \xi.
\end{eqnarray}
\end{subequations}
Using Efimov's radial law, it is derived that
\begin{equation}
\label{eq:DescrEfimovFreeSpace}
H = \kappa_* e^{\Delta(\xi)/2 s_0} e^{-n \pi/s_0},
\end{equation}
for a function $\Delta: [-\pi,-\pi/4] \to \mathbb{R}$. Here $\kappa_*$ is the three-body parameter that defines the trimer energies exactly on resonance as $E^n_T=-e^{-2 n\pi/s_0}\hbar^2 \kappa^2_*/m$, with $m$ the mass of a single atom, and is a result of the  short-range behavior of three particles being close together. The combinations of $K$ and $\xi$ that yield bound states describe a family of curves that can be mapped on top of each other by multiplying the distance of the curves to the origin by a factor $e^{\pi/s_0}$, see Fig.~\ref{fi:efimovschematic}. 

In this paper, we extend the description of Ref.~\cite{Braaten&Hammer06} and  show how by introducing the harmonic oscillator length as a new parameter, and by using spherical coordinates in parameter space, the scaling behavior in free space very naturally generalizes to a situation with three particles in a harmonic trap. In particular, the curves that are shown in Fig.~\ref{fi:efimovschematic} will be replaced by surfaces that are mapped on top of each other. In the derivation, we use an approach slightly different from earlier work, that calculates the equivalent of the function $\Delta$ by solving an ordinary differential equation. Additionally, we also include the coupling terms, we investigate the stability of the trimers in the lattice, and we study the non-universal corrections coming from the effective range parameter.

\section{Framework}

We consider a system of three bosons with pair interactions trapped in a harmonic potential. The corresponding Hamiltonian is

\begin{equation}
H = - \frac{1}{2} \sum_i \Delta_i + \frac{1}{2 \, \aho^4} \sum_i r_i^2 + \sum_i V_i,
\end{equation}
where $\aho= \sqrt{\hbar/m \omega}$ is the harmonic oscillator length,  $\omega$ the trap frequency, and $V_i$ is the interaction between the particles $j,k$, with $(i,j,k)$ a cyclic permutation of $(1,2,3)$. For convenience, the Hamiltonian is scaled by a factor $m/\hbar^2$. Throughout this paper, we will denote the eigenvalues of $H$ by $E$, and the corresponding trimer energies by $E_T=\hbar^2 E/m$. To find the bound state energies, we closely follow the method described in Ref. \cite{Fedorov93} and Ref. \cite{Braaten&Hammer06}. We denote the permutation operator corresponding to one cyclic permutation of $(1,2,3)$ by $P$ and transform to the Faddeev operator
\begin{equation}
F = - \frac{1}{2} \sum_i \Delta_i + \frac{1}{2 \, \aho^4}\sum_i r_i^2 + V(I + P + P^\dagger).
\end{equation}
If $\Psi$ is an eigenfunction of $F$, $(I + P + P^\dagger) \Psi$ is an eigenfunction of $H$, provided it is nonzero.  In case $(I+P + P^\dagger) \Psi = 0$, $\Psi$ is called a spurious solution. We separate the center of mass motion, and consider the relative system in hyperspherical coordinates. For that, we introduce Jacobi-coordinates according to
\begin{eqnarray}
\ve{r}_{i,jk} &=& \sqrt{\tfrac{2}{3}} (\ve{r}_i - \tfrac{1}{2}(\ve{r}_j + \ve{r}_k)),\\
\ve{r}_{jk} &=& \sqrt{\tfrac{1}{2}} (\ve{r}_j- \ve{r}_k).
\end{eqnarray}
The hyperradius $R$ and the hyperangle $\alpha$ are defined by
\begin{eqnarray}
R^2 &=& \frac{1}{3}(r_{ij}^2+r_{jk}^2+r_{ki}^2)=r_{i,jk}^2 + r_{jk}^2,\\
r_{i,jk}^2 &=& R \cos \alpha_i,\\
r_{jk}^2 &=& R \sin \alpha_i.
\end{eqnarray}
The remaining hyperspherical coordinates are given by the unit vectors $\hat{\ve{r}}_{i,jk}$ and $\hat{\ve{r}}_{jk}$.

We restrict ourselves to states with zero total angular momentum. The advantage of considering the Faddeev rather than the Schr\"{o}dinger operator is that subsystem angular momentum comes in only at second order in the short-range potential \cite{Nielsen01}. This motivates us to neglect the subsystem angular momentum, by averaging over the Jacobi coordinates $\hat{\ve{r}}_{i,jk}$ and $\hat{\ve{r}}_{jk}$. In this approximation, the Schr\"{o}dinger wavefunction $\Psi(\ve{r}_1,\ve{r}_2,\ve{r}_3)$ is given in terms of reduced Faddeev wavefunctions $\psi(R,\alpha)$:

\begin{equation}
 \Psi(\ve{r}_1,\ve{r}_2,\ve{r}_3)=\psi(R,\alpha_1)+\psi(R,\alpha_2)+\psi(R,\alpha_3).
\end{equation}
In the differential Faddeev equation~\cite{Braaten&Hammer06} we then make an adiabatic hyperspherical expansion

\begin{equation}
\label{eq:expansionhypang}
\psi(R,\alpha) = \frac{1}{2 \pi R^{5/2}\sin(2 \alpha)}\sum_n f_n(R) \phi_n(R,\alpha).
\end{equation}

For every value $R$, which is treated as a parameter, the functions $\phi_n(R,\alpha)$ of hyperangle $\alpha$ form a complete set, and are solutions to the eigenvalue equation on $L^2((0,\pi/2))$
\begin{equation}
\label{eq:eqhypwf}
\left[-\frac{\partial^2}{\partial \alpha^2} - \lambda_n(R)\right]\phi_n(R,\alpha) = - 2 R^2 V(\sqrt{2} R \sin \alpha)  \left[ \phi_n(R,\alpha) + \frac{4}{\sqrt{3}} \int_{|\frac{1}{3}\pi-\alpha|}^{\frac{1}{2}\pi - |\frac{1}{6}\pi-\alpha|} \phi_n(R,\alpha') d \alpha' \right],
\end{equation}
with Dirichlet boundary conditions $\phi(R,0) = \phi(R,\pi/2) = 0$. The eigenvalues $\lambda_n(R)$ depend parametrically on $R$. The set of hyperangular functions $\phi_n(R,\alpha)$ are generally not orthogonal functions in $\alpha$. Therefore, we define
\begin{equation}
\label{eq:G}
G_{nm}(R) = \int_0^{\frac{1}{2}\pi} \phi_n^*(R,\alpha)\phi_m(R,\alpha) d \alpha,
\end{equation}
which allows us to derive the following coupled set of differential equations for the functions $f_n$

\begin{multline}
\label{eq:coupledfn} \left[\frac{1}{2}\left(-\frac{\partial^2}{\partial R^2}+\frac{15}{4R^2}\right) + \frac{1}{2}\frac{1}{\aho^4}R^2 + \frac{\lambda_n(R) - 4}{2 R^2} \right] f_n(R) +\\+ \sum_m \left[2 P_{nm}(R) \frac{\partial}{\partial R} + Q_{nm}(R)\right]f_m(R) = E f_n(R).
\end{multline}
The coupling terms in these equations are defined by
\begin{subequations}
\label{eq:couplingterms}
\begin{eqnarray}
P_{nm}(R) &=& - \frac{1}{2} \sum_k G_{nk}^{-1}(R) \int_0^{\frac{1}{2}\pi}  \phi_k^*(R,\alpha) \frac{\partial}{\partial R}\phi_m(R,\alpha) d \alpha,\\
Q_{nm}(R) &=& - \frac{1}{2} \sum_k G_{nk}^{-1}(R) \int_0^{\frac{1}{2}\pi}  \phi_k^*(R,\alpha) \frac{\partial^2}{\partial R^2}\phi_m(R,\alpha) d \alpha.
\end{eqnarray}
\end{subequations}

This framework is very similar to the work of Refs.~\cite{Fedorov93,Braaten&Hammer06}, and the harmonic oscillator term is taken into account as in Ref.~\cite{Jonsell02}. For now, we neglect the coupling terms conform the adiabatic hyperspherical approximation. However, later on we wil study the coupling between different channels more carefully, as the harmonic oscillator gives rise to degeneracies between the Efimov-like solutions in the zeroth channel and the continuum-like solutions of the first and higher channels.\\
\newpage
\indent {\bf Hyperangular problem}\\

The effect of the interaction potential $V$ is now encoded in the eigenfunctions $\lambda_n(R)$. Parametric solutions for $\lambda_n$ can be found by fixing the parameter $R$, while solving for the angular problem. The functions $\lambda_n(R)$ will then serve as input for the radial equation (\ref{eq:coupledfn}), and when the coupling terms are ignored, solutions are obtained by solving an ordinary differential equation (ODE).

To find an approximation of $\lambda_n$, one can study solutions $\psi(R,\alpha)$ of Eq.~(\ref{eq:eqhypwf}) for small values of $\alpha$ and values of $\alpha$ close enough to $\pi/2$. Matching of the solutions at intermediate values of $\alpha$ then yields an approximation of $\lambda_n$. This procedure has been described in Ref.~\cite{Fedorov93}. The equation found in this manner is
\begin{equation}
\cos\left(\lambda_n^{1/2}\tfrac{\pi}{2}\right)-\frac{8}{\sqrt{3}}\lambda_n^{-1/2}\sin\left(\lambda_n^{1/2} \tfrac{\pi}{6}\right) = - \sqrt{2}\lambda_n^{-1/2} \sin\left(\lambda_n^{1/2}\tfrac{\pi}{2}\right) R \; k_R \cot(\delta(k_R)),
\label{eq:approxlambda} 
\end{equation}
where $k_R=\sqrt{\lambda_n/2R^2}$. Note that $\lambda_n\equiv16$ always solves this equation, but it gives rise to a spurious solution. The right hand side of equation (\ref{eq:approxlambda}) can be conveniently expanded using the effective range approximation for the s-wave scattering phase shift of the two-body problem
\begin{equation}
k_R \cot \delta(k_R) =  - \frac{1}{a} + \frac{1}{2} r_s k_R^2 + O(k_R^4),
\label{eq:EffRangeExp}
\end{equation}
with $r_s$ the effective range parameter. When this parameter is neglected, we find the approximating equation for $\lambda_n = \lambda_n(R/a)$ that is at the basis of the derivation in Ref.~\cite{Fedorov93}:
\begin{equation}
\label{eq:lambdaonlya} \cos\left(\lambda_n^{1/2}\tfrac{\pi}{2}\right)-\frac{8}{\sqrt{3}}\lambda_n^{-1/2}\sin\left(\lambda_n^{1/2} \tfrac{\pi}{6}\right) = \\  \sqrt{2}\lambda_n^{-1/2}(R) \sin\left(\lambda_n^{1/2}\tfrac{\pi}{2}\right) \frac{R}{a}.
\end{equation}
By taking into account the effective range as well, one obtains an approximation that is more accurate with respect to the two-body physics near narrow Feshbach resonances. This has been proposed in Ref.~\cite{Fedorov01} as a regularization procedure and has been readdressed in Ref.~\cite{Platter09}. We will come back to this issue in Section \ref{sec:effrange}.\\

\indent {\bf Hyperradial problem}\\

With the approximation of the channel eigenvalues at hand, we may turn to the hyperradial problem. In the adiabatic hyperspherical approximation, we ignore the coupling between the different hyperradial channels and only take the scattering length into account in the approximation of $\lambda_n$. In this setting, the Efimov effect was derived in Ref.~\cite{Fedorov93}, and repeating it in this context allows for a comparison in the limit of $a_{\rm ho} \rightarrow \infty$. The uncoupled system of equations is given by

\begin{equation}
\label{eq:hyperradial}
\Big[-\frac{\partial^2}{\partial R^2} + \frac{\lambda_n(R)-\tfrac{1}{4}}{R^2} \\ +\frac{R^2}{a_\text{ho}^4} \Big]f_n(R)= 2 E f_n(R),
\end{equation} 
on $(R_0, \infty)$, with a boundary condition 
\begin{equation}
f_n(R) \to 0 , \quad  R\to \infty,
\end{equation}
and an appropriate boundary condition on the edge of the inner region $R_0$ where the short-range potentials become important, which is given in terms of the three-body parameter $\kappa_*$. The exact boundary condition will be specified later.

We first concentrate on the zeroth channel, i.e.~$n=0$, which is the only channel that has an attractive potential, and therefore the only channel in which the Efimov effect is found. We introduce spherical coordinates $(T,\vartheta,\xi)$ according to

\begin{subequations}
\label{eq:SpherCor}
\begin{eqnarray}
K &=& T \sin \vartheta \sin \xi,\\
\frac{1}{a} &=& T \sin \vartheta \cos \xi,\\
\frac{1}{a_\mathrm{ho}} &=& T \cos \vartheta.
\end{eqnarray}
\end{subequations}
and write down the differential equation for $f(\sigma)=f_0(\sigma/T)$ as

\label{subeq:problemforf}
\begin{equation}
\label{eq:diffeqforf} 
\Big[-\frac{\partial^2}{\partial \sigma^2} + \frac{\lambda_0(\sigma \sin \vartheta \cos \xi)-\tfrac{1}{4}}{\sigma^2} + \cos^4 \vartheta \; \sigma^2 \Big]f(\sigma)  = 2\sgn{(\xi)}\sin^2 \vartheta \sin^2 \xi f(\sigma),
\end{equation}
with long-range boundary condition
\begin{equation}
\label{eq:bclargesigma}
f(\sigma) \to 0, \quad \sigma \to \infty.
\end{equation}
For small values of $\sigma$, the model does not provide us with a natural boundary condition. Solutions of Eq.~(\ref{eq:diffeqforf}) have a log-periodic behavior as $\sigma \rightarrow 0$, given by $f(\sigma) \sim \sqrt \sigma \sin(s_0 \log(\kappa \sigma/T) + \alpha_0)$ for some value of $\kappa$. Here $s_0 = \sqrt{- \lambda_0(0)}$, and 
\begin{equation}
\alpha_0 = - \frac{1}{2} s_0 \log 2 - \frac{1}{2}
\mathrm{Arg}\frac{{\rm\Gamma}(1 + i s_0)}{{\rm\Gamma}(1 - i s_0)}.
\end{equation}

Fortunately, the real short-range potentials $V_i$ provide us with a natural boundary condition at $R_0$ encoded by $\kappa=\kappa_*$, where $\kappa_*$ is the three-body parameter that accounts for the short-range physics, which was mentioned already in the introduction. This gives rise to a short-range boundary condition at $R_0$:
\begin{equation}
\label{eq:bcsmallsigma}
\frac{f'(\sigma_0)}{f(\sigma_0)} = \frac{1}{2 \sigma_0} + \frac{s_0}{\sigma_0} \cot(s_0 \log(\kappa_* R_0) + \alpha_0).
\end{equation}

\begin{figure}[th]
\begin{center}
\includegraphics[width=8cm]{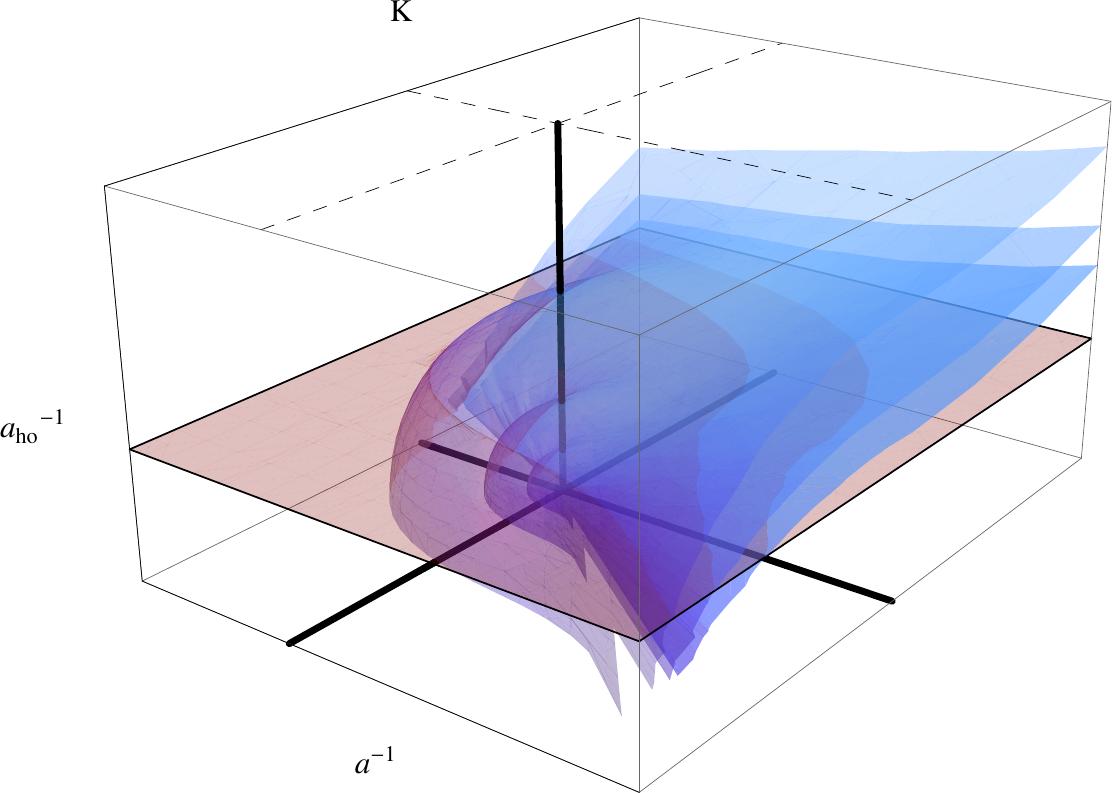}
\caption{Surfaces that represent trimer bound states in the lowest channel, as a function of $a^{-1}$, $K$ and $a_\text{ho}^{-1}$ (blue). Only three surfaces are shown, but in fact there is an infinite number of them, which all can be mapped on top of each other by multiplying the distance to the origin by powers of $e^{\pi/s_0}$. These surfaces are sliced by a plane (red transparant), for a given $a_\text{ho}$ that corresponds to the height of the plane. The intersection lines (red) represent bound state solutions for combinations of the scattering length $a$ and the wave vector $K$, for this value of  $a_\text{ho}$. The picture is not too scale: we used similar scaling relations as used in Fig.~\ref{fi:efimovschematic} in order to show the surfaces more clearly.}
\label{fi:boundstatesurfaces}
\end{center}
\end{figure}

We may determine for which parameter combinations a solution exists by a shooting method from large values of $\sigma$. A particular convenient method makes use of a certain phase variable $\phi$. One can set up an equation for $\phi$ that is much alike the variable phase equation, but is adapted to fit the boundary condition at short distances very well. Henceforth, it will play a role similar to the function $\Delta$ as we mentioned in the introduction. We introduce the transformation
\begin{equation}
f'(\sigma) = \big(g(\sigma) + h(\sigma) \cot( \phi(\sigma) + \psi(\sigma) )\big) f(\sigma).
\end{equation}
The functions $g$, $h>0$ and $\psi$ may be chosen to our convenience. Continuity properties in the parameters are easier to derive considering $\phi$ than when one considers $f$ directly. A particular choice
\begin{eqnarray}
g &=& \frac{1}{2 \sigma}, \\
h &=& \frac{s_0}{\sigma}, \\
\psi &=& s_0 \log \sigma + \alpha_0,
\end{eqnarray}
is highly compatible with the boundary condition (\ref{eq:bcsmallsigma}). The equation for $\phi$ becomes 
\begin{equation}
\label{eq:forphief}
s_0 \phi' + \Big(\frac{\lambda_0(\sigma \sin \vartheta \cos\xi)+s_0^2}{\sigma} -2 \sigma \sgn(\xi) \sin^2 \vartheta \sin^2 \xi  + \cos^4 \vartheta \sigma^3\Big)\sin^2{(s_0 \log{\sigma} +\alpha_0 + \phi(\sigma))} = 0,
\end{equation}
which is a nonlinear first order differential equation. Given values for $\xi$ and $\vartheta$, we denote by $\phi(\sigma;\xi,\vartheta)$ the solution to (\ref{eq:forphief}) that satisfies $\phi(\sigma) + s_0 \log \sigma + \alpha_0 \uparrow 0$. With this definition, the problem (\ref{subeq:problemforf}) has a solution if and only if
\begin{equation}
\phi(T R_0 ; \xi, \vartheta) = s_0 \log (\kappa_*/T) \;\mathrm{mod}\; \pi.
\end{equation}
Consequently, eigensolutions only exist if there exists an $n \in \mathbb{Z}$ such that
\begin{equation}
\label{eq:maintheorem}
T = \kappa_* e^{-\phi(T R_0; \xi, \vartheta)/s_0}e^{-n\pi/s_0}.
\end{equation}
We would like to emphasize here the similarity with the description of the Efimov effect in free space given by equation (\ref{eq:DescrEfimovFreeSpace}). Using the asymptotic behavior of $\lambda_0$ for small arguments, one can easily show that $\phi(\sigma)$ converges to a finite value for $\sigma \downarrow 0$. The function $\phi(0;\xi,\vartheta)$ can be shown to be continuous over the region. For $\vartheta=0$ and $\xi \in [-\pi,-\pi/4]$, the Efimov effect in free space is retained. Hence, $\phi(0;\xi,0)$ is an approximation of $-\Delta(\xi)/2$.\\

\indent {\bf Trimer solutions in a harmonic potential}\\
\label{se:resultssimplemodel}

\begin{figure}
\begin{center}
\includegraphics[width=8cm]{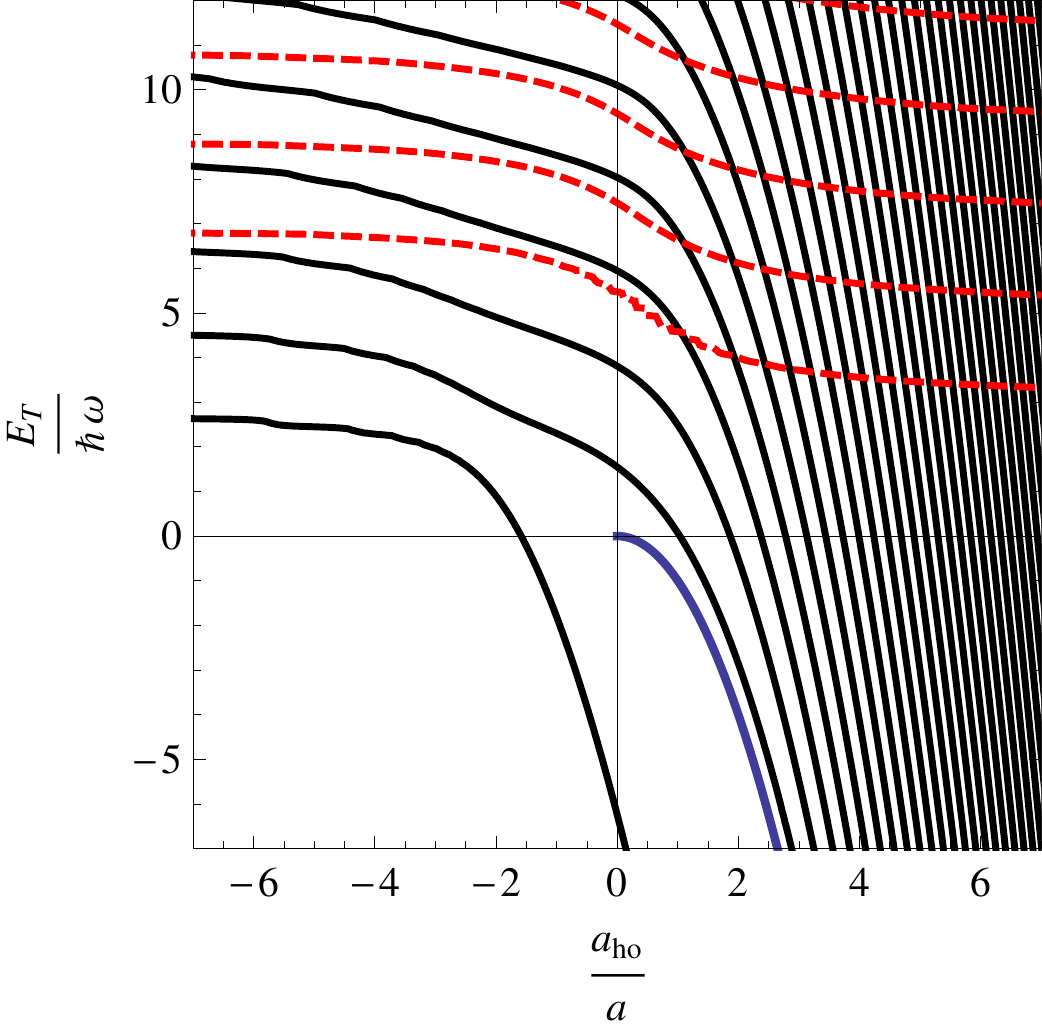}
\caption{The trimer bound state energies of cesium in the zeroth and first hyperspherical channels. The result is obtained for a trap frequency $\omega = 1000$ s$^{-1}$. A deeper bound state, with much lower energy, can not be shown in this graph. Also indicated is the dimer binding energy (solid blue line)}
\label{fi:channelscombined}
\end{center}
\end{figure}

The trimer solutions are most easily represented by surfaces in three-dimensional $(a^{-1},K,a_\text{ho}^{-1})$ space, as defined in Eq.~(\ref{eq:SpherCor}). In Fig.~\ref{fi:boundstatesurfaces} such surfaces are shown for the lowest channel, which can be mapped on top of each other by multiplication of the distance to the origin by powers of $e^{\pi/s_0}$.  
Fig.~\ref{fi:boundstatesurfaces} can be regarded as a generalization of the two-dimensional Fig.~\ref{fi:efimovschematic}. When these surfaces are sliced with a plane corresponding to a fixed harmonic oscillator length $a_\text{ho}$, we obtain curves that express wave vectors $K$ in terms of the scattering length $a$, and an example can be seen in Fig.~\ref{fi:channelscombined}. When taking the limit of $a_\text{ho} \to \infty$, the Efimov effect in free space is obtained as depicted in Fig.~\ref{fi:efimovschematic}. In the region $\xi \in [-\pi,-\pi/4]$, the intersections of the surfaces will tend to the curves that describe the Efimov states. In the region $\xi \in (-\pi/4,\pi)$, the energy levels come closer and closer together, expressing the fact that in the limit, this region is filled with continuum states. The intersection of the surfaces with the $a^{-1} = 0$ plane precisely corresponds to the exact results obtained by Jonsell {\it et al.} and Werner~\cite{Jonsell02,Werner08}.

As an example, we show in Fig.~\ref{fi:channelscombined} the trimer binding energies of cesium as a function of scattering length with a fixed oscillator length, for the lowest and next to lowest hyperspherical channels. The parameters are $\kappa_* = 1.93 \times 10^{-4} a_0^{-1}$~\cite{Kraemer06}, with $a_0$ the Bohr radius, $R_0 = 100 a_0$, and $a_\text{ho} = 13000a_0$, which corresponds to a trap frequency of approximately $1000$ s$^{-1}$. 

It is interesting to note the change of energy scales in this figure. Below the thresholds of the trimer and atom-dimer continuum (cf.~Fig.~\ref{fi:efimovschematic}), which would exist for the free-space Efimov effect, the bound states show the characteristic scaling law of Eq.~\ref{efimovscaling}. Above these thresholds, the bound states show a spacing typical for the harmonic oscillator potential which is approximately $2\hbar\omega$. This change of exponential scaling to linear scaling of the trimer energies makes that the accumulation of bound states in the point of zero energy and infinite scattering length disappears. Effectively, the Efimov trimers are "pushed" out of the trimer region through the thresholds and aquire a strong harmonic oscillator state character.

\begin{figure}
\begin{center}
\includegraphics[width=8cm]{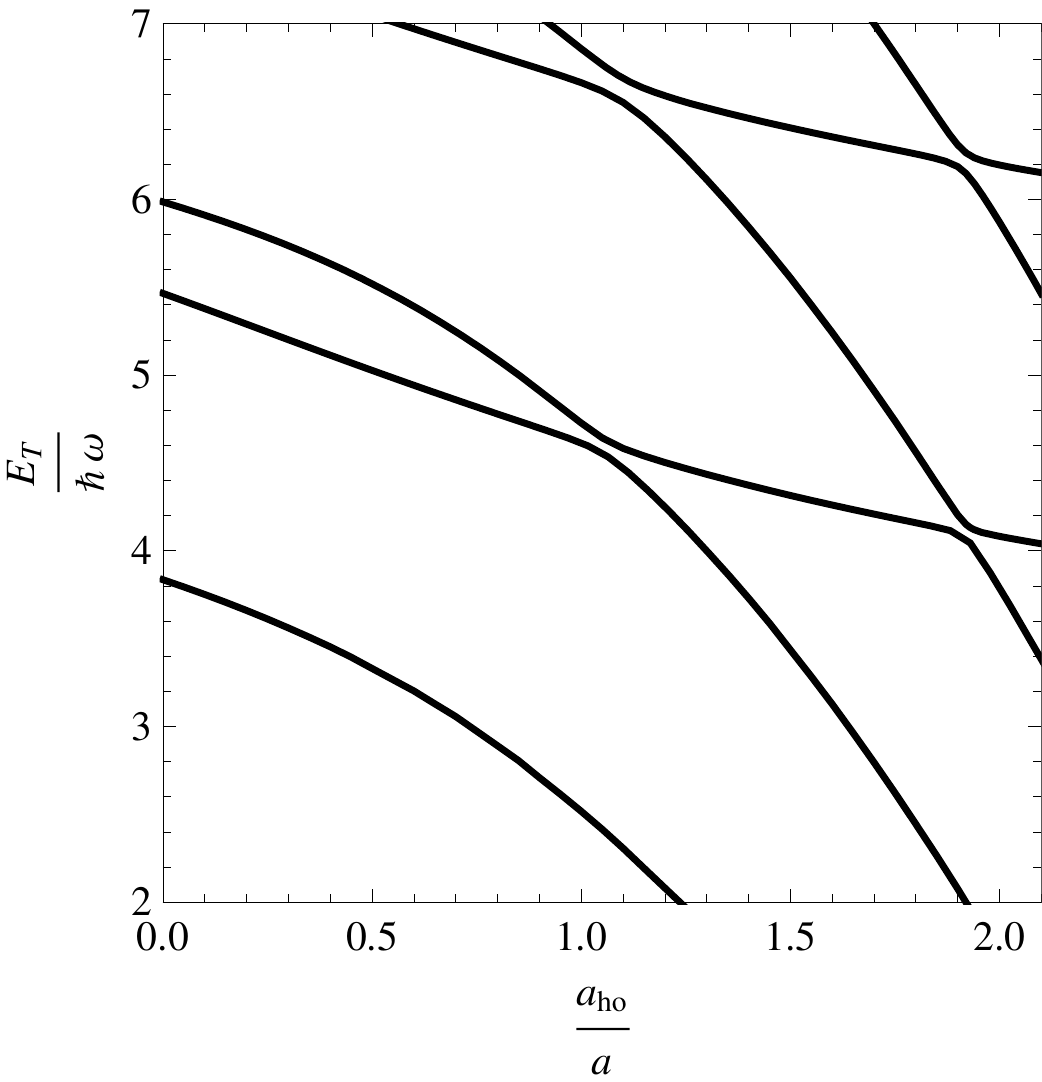}
\caption{Crossings in Fig.~\ref{fi:channelscombined} are replaced by avoided crossings when the coupling between the zeroth and first hyperspherical channels is taken into account.}
\label{fi:avoidedcrossings}
\end{center}
\end{figure}

There is also a clear difference between the zeroth and first hyperspherical channels. In this respect, it is interesting to note the different limiting behavior of the solutions for small positive and negative scattering length. For small and negative scattering length, there is no weakly-bound dimer state in the interatomic potential, and the solutions have the character of three weakly-interacting atoms in a harmonic oscillator potential. For small and positive scattering length, there is a weakly-bound dimer state, and for the zeroth hyperspherical channel the solutions have the character of a weakly-interacting dimer and an atom in a harmonic oscillator potential. These regimes are connected through solutions which have a strong Efimov character. However, for the first hyperspherical channel the situation is different. Here the solutions  behave more as those of two particles undergoing a Feshbach resonance~\cite{Busch98,Stoferle06}, which as a result interpolate between two different harmonic oscillator levels.

It should be possible to investigate this change of character experimentally, by using similar techniques as have been done by St\"oferle {\it et al.}~\cite{Stoferle06}. Here the creation of dimers was investigated in an optical lattice while sweeping the magnetic field through a Feshbach resonance. For the trimers, such adiabatic sweeps through trimer solutions should allow for a creation of states with a character changing between harmonic oscillator-, Efimov- and dimer+atom- type solutions. 

So far, we have applied the adiabatic hyperspherical approximation, in which we completely ignored the coupling between the various hyperspherical channels. While this is generally a good approximation, we see from Fig.~\ref{fi:channelscombined} that there are degeneracies between the zeroth and first channel solutions, which should be lifted when the coupling is taking into account. In order to demonstrate this, we introduce an approximation of the coupling. We calculate the coupling terms $P_{mn}$ and $Q_{mn}$ of Eq.~(\ref{eq:couplingterms}) using first order approximations of the functions $\phi_n$, and find in this way combined solutions which involve both channels. In free-space, the coupling between these two hyperspherical channels is directly related to the process of three-body recombination, where three colliding particles relax into a weakly-bound dimer and a third particle, with a surplus of kinetic energy~\cite{Braaten&Hammer06}. Interestingly, for our lattice solutions we observe that this coupling transforms the crossings in Fig.~\ref{fi:channelscombined} into avoided crossings, as shown in Fig.~\ref{fi:avoidedcrossings}. This could be experimentally interesting, since using magnetic field sweeps would allow for adiabatic transitions from one hyperspherical channel to an other.

\section{Stability of trimers in a harmonic potential}

\begin{figure}
\begin{center}
\includegraphics[width=10cm]{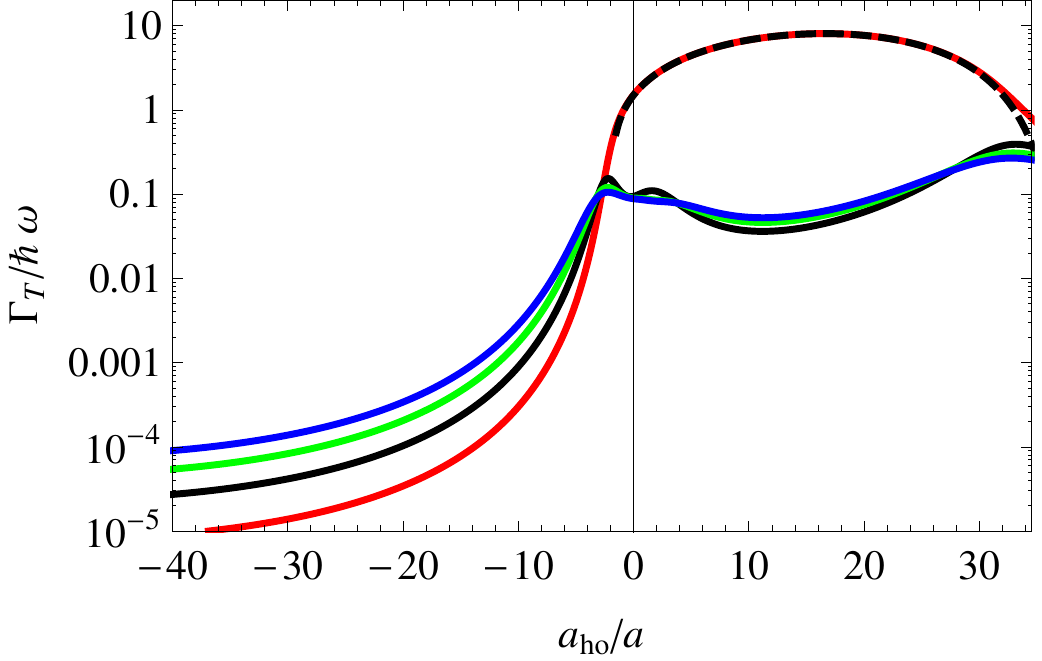}
\caption{The width $\Gamma_T$ of cesium Efimov states in a trap, with $\omega=1000$ s$^{-1}$ and $\eta_* = 0.06$, for the first (red), second (black), third (green) and fourth (blue) trimer states depicted in Fig. \ref{fi:channelscombined}. Also, the width of a free Efimov trimer corresponding to the first state is shown (black dashed).} 
\label{fi:stabplot}
\end{center}
\end{figure}

The change of character of the trimer solutions, which was discussed above, will have a profound impact on the stability of these states. So far, we implicitly assumed that the trimers in the harmonic oscillator potential are stable. However, when the three particles are close together the trimer can dissociate into a deeply-bound dimer and a third atom, with large kinetic energy. Here, we investigate the stability of the trimers by applying the method of Braaten and Hammer~\cite{Braaten&Hammer06}, that regards the trimers as scattering resonances in a dimer-atom scattering process. In this respect, the trimer energy $E_T$ acquires a width $\Gamma_T$, where the latter can be used to calculate the lifetime $\hbar/\Gamma_T$ of the trimers. In this method, an inelasticity parameter $\eta_*$ is introduced to account for the decay into 
various deeply bound states. which can be regarded as the imaginary part of a complex phase factor that describes the reflection of hyperradial waves in the inner region. Effectively, this comes down to changing the boundary condition at $R_0$ to 
\begin{equation}
\frac{f'(R_0)}{f(R_0)} = \frac{1}{2 R_0} + \frac{s_0}{R_0} \cot\left(s_0 \log (\kappa_* R_0) + \alpha_0 + i \eta_* \right).
\end{equation}
For given values for the scattering length and the harmonic oscillator length, we are then able to find the associated complex eigenvalues

\begin{equation}
E = \frac{m}{\hbar^2}\left(E_T - i \frac{\Gamma_T}{2}\right),
\end{equation}
where the trimer energy $E_T<0$ and the width $\Gamma_T > 0$. 

As an example, we calculated the linewidths or cesium, with a trapping frequency $\omega = 1000$ s$^{-1}$, $\kappa_* = 1.93\times 10^{-4}a_0$, and $\eta_* = 0.06$~\cite{Kraemer06}. The short-range boundary condition was implemented at a value $R_0=100 a_0$. For some of the Efimov trimers shown in Fig. \ref{fi:channelscombined}, the linewidths are shown in Fig. \ref{fi:stabplot}. The change of character of the trimers can be clearly seen from this graph: for values  $a/a_{\rm ho}\lesssim-2$ the states have a harmonic oscillator character, and the widths are relatively small. However, the Efimov effect shows up via a strong increase of the widths towards $a/a_{\rm ho} \gtrsim -2$, which is the parameter region where in free-space a new Efimov state would enter the tree-body system. It is also the region where a local peak in the three-body recombination rate would occur (trimer resonance). Here, we see that at this point the width of the lowest trimer keeps increasing, while for the higher trimer states the width saturates and displays an oscilating behavior. This is not surprising, since the latter trimers are more similar to three-particle continuum states. For comparison, the linewidth of the free-space Efimov trimer compared to the lowest lattice trimer state is shown in the region where it exists. These lines overlap mostly, and can be more than two orders of magnitude above the widths for the higher trimer states. At the point where the trimer energies approach the atom-dimer threshold, around $a/a_{\rm ho}=34$, the widths go down again to the level of the higher trimer states. The width of the lattice trimer is there somewhat larger than the free-space trimer, probably due to the additional confinement of the weakly-bound dimer state caused by the lattice.

While the Efimov effect shows up in the excited states by a clear deviation from the harmonic oscillator levels, these states benefit from a much smaller linewidth than the deeply bound Efimov trimers, and they should therefore be easier to study experimentally. Also the lowest trimer energy, which most closely follows a free-space Efimov state, could be easier to study in experiment. The trimer could first be prepared in the lowest harmonic oscillator level for $a/a_{\rm ho}\ll -1$, and then the scattering lenght can be changed via a magnetic field sweep to the region $a/a_{\rm ho}\approx-2$, where the decay is still manageable.  

\section{Effective range corrections} \label{sec:effrange}

In order to improve the approximation of the channel eigenvalues $\lambda_n$ arising in (\ref{eq:eqhypwf}), one might want to include the influence of the effective range $r_s$. This can be done by making the so-called effective range expansion, i.e.~using the expansion in Eq.~(\ref{eq:EffRangeExp}) up to second order. In literature there have been various instances in which the effective range is taken into consideration in the determination of the three-particle spectrum \cite{Petrov04,Gogolin08}. 

Having an additional interaction parameter at hand, raises the question whether in this regime Efimov physics in ultracold gases can still be universal. On the experimental side, there are presently also some conflicting observations. In order to understand these issues, we first need to define what we mean by the universal Efimov regime. In the above, we have already seen that besides the scattering length, we need at least one additional parameter that fixes at least one of the Efimov trimer energies. The other ones can then be derived from this position. This information is provided via the three-body parameter $\kappa_*$, which can be regarded as a boundary condition for the interaction potentials in the regime where the three particles are very close together. It is related to the energy of the lowest Efimov trimer via $E_T=-\hbar^2 \kappa_*^2/m$ in the infinite scattering length limit, and with this only additional parameter the Efimov physics is universal. 

\begin{figure}
  \center
  \includegraphics[width=7cm]{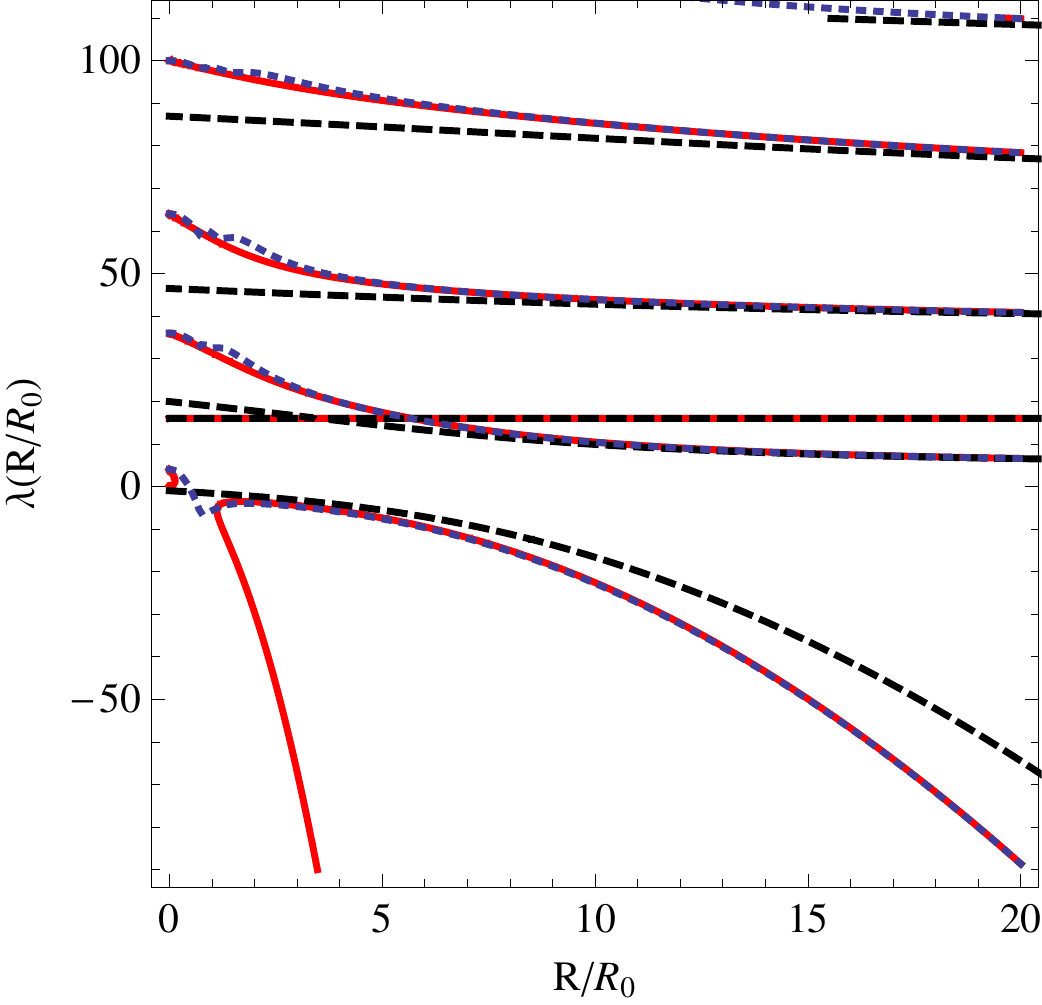}\\
  \caption{Eigenvalues $\lambda$ of $\mathcal{F}^R$ depending on $R/R_0$, as calculated in three different ways. The blue lines represent the approximations of the eigenvalues by a truncated Fourier expansion. The black lines and red lines are the approximations obtained by (\ref{eq:lambdaonlya}) and (\ref{eq:lambdawitheffrange}) respectively. We take for the square well parameters $V_0R_0^2=-3.2$. It can be clearly seen that the effective range gives a much better asymptotic behavior. }\label{fi:comparisoneffrange}
\end{figure}

However, some experiments on three-body recombination do not imply universality. While the experiments at Rice University did show the universal connection between Efimov states in $^7$Li at one side of the resonance~\cite{Pollack09}, the universality did not extend to the other side of the resonance. The Rice group was not alone in this observation: also in Florence~\cite{Zaccanti09} on $^{39}$K, in Innsbruck~\cite{Knoop09} on $^{133}$Cs, and in Tokyo~\cite{Nakajima10} on $^{6}$Li this non-universality was observed. On the other hand, in three other experiments universality across the resonance was observed: in Bar-Ilan it was found that universality does extend across a Feshbach resonance in $^7$Li, and even that the short range physics encapsulated in $\kappa_*$ is independent of the nuclear spin configuration~\cite{gross10,gross_specialtopic}. Furthermore universality was also observed in a three-component fermionic $^6$Li gas in Heidelberg~\cite{Ottenstein08} and in State College~\cite{Huckans09}.

Now the question arises whether the effective range paramer $r_s$ can explain, in addition to the three-body parameter $\kappa_*$, the non-universal behavior of some of the above-mentioned experiments. However, it might also be that in a certain regime, the effective range will take over the role of the three-body parameter due to its regularising effect of the radial wavefunctions~\cite{Petrov04,Gogolin08}. And recently, it has been shown that small experimental deviations from universality ~\cite{gross10} can be explained by effective range corrections using an effective field
theory~\cite{Ji10}.

As it has been shown by Fedorov and Jensen \cite{Fedorov01} and more recently by Platter {\it et al.} \cite{Platter09}, the introduction of the effective range in the approximate equation (\ref{eq:approxlambda}) for the eigenvalues $\lambda_n$ yields the relation

\begin{equation}
\label{eq:lambdawitheffrange} \cos\left(\lambda_n^{1/2}\tfrac{\pi}{2}\right)-\frac{8}{\sqrt{3}}\lambda_n^{-1/2}\sin\left(\lambda_n^{1/2} \tfrac{\pi}{6}\right) = - \sqrt{2}\lambda_n^{-1/2} \sin\left(\lambda_n^{1/2}\tfrac{\pi}{2}\right) \; \left( -\frac{R}{a} + \frac{1}{2}r_s \frac{\lambda_n}{2R}\right).
\end{equation}

To get an idea on the validity of the approximation (\ref{eq:lambdawitheffrange}), we have compared the results with the eigenvalues obtained by a truncated Fourier expansion for a square well potential with height $V_0$ and width $R_0$, such that $V_0 R_0^2 = -3.2$.
This expansion is very convenient when one considers the operator $\mathcal{F}^R$, which depends on the  hyperradius $R$ and is defined via the eigenvalue equation (\ref{eq:eqhypwf}) as follows
\begin{equation}
 \mathcal{F}^R \phi_n(R,\alpha) = -\frac{\partial^2}{\partial \alpha^2} \phi_n(R,\alpha) + 2 R^2 V(\sqrt{2} R \sin \alpha)  \left[ \phi_n(R,\alpha) + \frac{4}{\sqrt{3}} \int_{|\frac{1}{3}\pi-\alpha|}^{\frac{1}{2}\pi - |\frac{1}{6}\pi-\alpha|} \phi_n(R,\alpha') d \alpha' \right].
\end{equation}
The reason for this convenience is that in the basis of Fourier expansion functions
\begin{equation}
\xi_n(\alpha) = \frac{2}{\sqrt{\pi}} \sin(2 n \alpha),
\end{equation}
both the differential operator and the integral operator are diagonal. The matrix elements are then given by
\begin{equation}
(\mathcal{F}^R)_{mn} = 4 n^2 \delta_{mn} + 2 R^2\left(1+ \frac{4}{\sqrt{3}} \frac{1}{n} \sin(2 n \tfrac{\pi}{3})\right) \mathcal{V}_{mn}(R),
\end{equation}
where
\begin{equation}
\mathcal{V}_{mn}(R) = \frac{4}{\pi} \int_0^{\pi/2} V(\sqrt{2}R \sin \alpha)\sin(2 n \alpha) \sin(2 m \alpha) d \alpha.
\end{equation}
In Fig.~\ref{fi:comparisoneffrange}, we compare the numerically calculated eigenvalues $\lambda_n(R)$ of $\mathcal{F}^R$ with results obtained from Eq.~(\ref{eq:lambdaonlya}), which depends on the scattering length only, and with the results from solving Eq.~(\ref{eq:lambdawitheffrange}), which depends both on scattering length and effective range. Here the values for $a$ and $r_s$ are taken from the effective range expansion of the square well potential. It can be seen that asymptotically, that is for large hyperradius, the inclusion of the effective range in the model yields results that are considerably better than if one only takes the scattering length into account. For relatively small values of the hyperradius, the predictions obtained by the Fourier expansion and this simple model start to deviate. Hence, we expect that for short hyperradii, additional details of the interaction potential become important, i.e.~higher order expansion coefficients beyond the effective range approximation have to be taken into account.

For $r_s<0$ and a fixed value for $a$, the lowest solution $\lambda_0$ to Eq.~(\ref{eq:lambdawitheffrange}) can be defined as a smooth function of $R$ for $R\in(0,\infty)$. Moreover, $\lambda_0(R) = O(R)$ as $R \to 0$. Therefore, if we use this approximation in the hyperradial equation, we can find \emph{regular} solutions instead of rapidly oscillating ones. This could, at least in principle, remove the need for a renormalization boundary condition in terms of $\kappa_*$ for $R \to 0$. In the absence of a harmonic oscillator, and in the limits of small energy and $r \equiv -r_s/a \to 0$, we will see that the solutions to the hyperradial equation are identical to those when the effective range is not taken into account, except that the value of $\kappa_*$ is determined by $r_s$ by boundary layer behavior for small values of the hyperradius. Such a relation between $\kappa_*$ and $r_s$ has been established before \cite{Petrov04,Gogolin08}.

\begin{figure}
\begin{center}
\includegraphics[width=8cm]{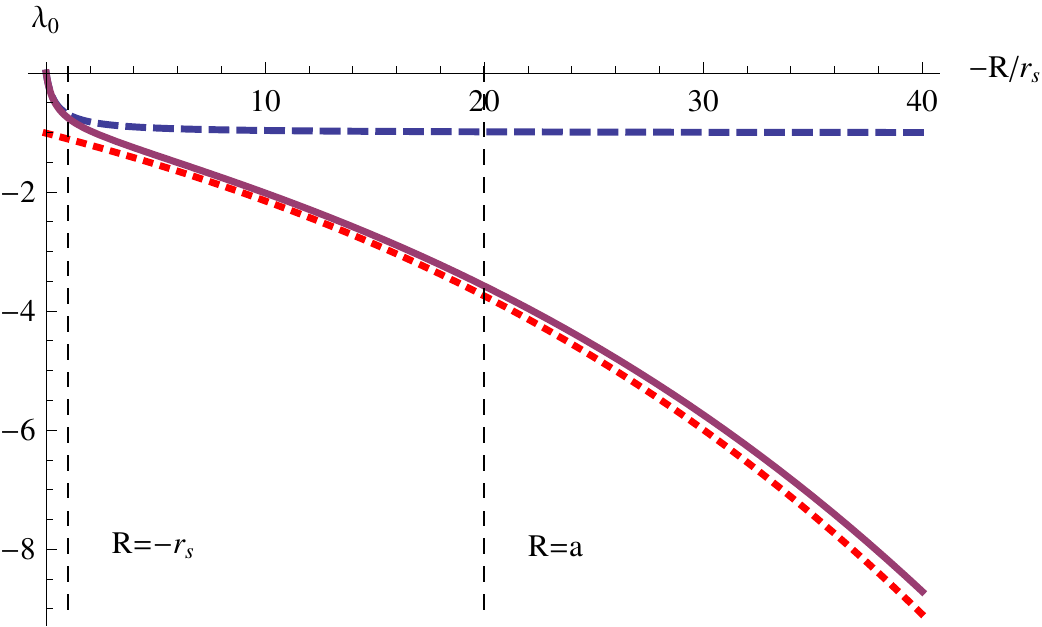}
\caption{The figure shows $\lambda_0$ (purple, solid) as a function of $-R/r_s$, for a fixed value of $a = -20 r_s$, with $r_s < 0$. For $R \gg - r_s$, the solution is very close to that obtained from (\ref{eq:lambdaonlya}) (red, dotted). For $R \lesssim -r_s$, there is a boundary layer, in which the solution resembles the solution $\mu$ of (\ref{eq:mu}) (blue, dashed).}
\label{fi:illlambda0}
\end{center}
\end{figure}

Let us, therefore, study the equation (\ref{eq:lambdawitheffrange}) more closely. We have plotted the solution $\lambda_0$ for the situation $a = - 20 r_s$ in Fig. \ref{fi:illlambda0}, to give a characteristic example. In the region $-r_s \ll R \lesssim a$, the value of $\lambda_0$ satisfying (\ref{eq:lambdawitheffrange}) is close to that obtained by solving (\ref{eq:lambdaonlya}). Similarly, when $R \ll -r_s$, the solution $\lambda_0$ to (\ref{eq:lambdawitheffrange}) will be close to $\mu(-R/r_s)$, where $\mu(-R/r_s)$ is the solution to

\begin{equation}
\label{eq:mu} \cos\left(\mu^{1/2}\tfrac{\pi}{2}\right)-\frac{8}{\sqrt{3}}\mu^{-1/2}\sin\left(\lambda_n^{1/2} \tfrac{\pi}{6}\right) = - \sqrt{2}\lambda_n^{-1/2} \sin\left(\mu^{1/2}\tfrac{\pi}{2}\right) \; \frac{1}{2}r_s \frac{\mu}{2R}.
\end{equation}
Around $R \sim -r_s$, there is a transition from a boundary layer behavior similar to $\mu$ to behavior obtained by ignoring the effective range (\ref{eq:lambdaonlya}). 

To find the low-energy behavior of the system, which is needed to find a proper relationship between $r_s$ and $\kappa_*$ \cite{Petrov04,Gogolin08}, we have to turn back to the hyperradial equation (\ref{eq:hyperradial}). Following from the above considerations, the solution $f_0$ that decays exponentially for large hyperradius is to leading order in $-r_s/a$, and for $R \gg -r_s$, equal to $f^{\text{out}}$, the exponentially decaying solution to the hyperradial equation obtained when the effective range is ignored. This outer solution $f^{\text{out}}$ will have log-periodic behavior as $R \to 0$ 
\begin{equation}
f^{\text{out}}(R) \sim \sqrt{R} \sin(s_0 \log(\kappa R) + \alpha_0),
\end{equation}
for some $\kappa$. The solution $f_0$ will be similar for $-r_s \ll R \ll a$, and will therefore for low enough energies also show log-periodic behavior. However, there is a boundary layer for $R \ll -r_s$, which makes that we can demand the solution $f_0(R)$ to be regular for $R \to 0$, that is, such that $f_0(R) \sim \sqrt{R}$ for $R \to 0$. This effectively determines the value of $\kappa$. To analyze the boundary layer, we introduce the new variable $\rho = -R /r_s$ and the function $\tilde{f}_0$, $\tilde{f}_0(\rho) = f_0(- \rho r_s)$. The hyperradial equation for $\tilde{f}_0$ reads 
\begin{equation}
\left[ - \frac{d^2}{d\rho^2} + \frac{\lambda_0(\rho) - 1/4}{\rho^2} + \frac{r_s^4}{a^4_\text{ho}}\rho^2 - 2 E r_s^2 \right] \tilde{f}_0(\rho) = 0,
\end{equation}
and we demand that $\tilde{f}_0(\rho)$ behaves like $\sqrt{\rho}$ for $\rho \to 0$. We consider now the case of a negligible harmonic oscillator potential, i.e. $a_{\rm ho}=\infty$, and take the low-energy limit $E r_s^2 \to 0$ and a diverging scattering length. The solution will then be equal to $f_0^{\text{in}}$, which satisfies
\begin{equation}
\left[ - \frac{d^2}{d\rho^2} + \frac{\mu(\rho) - 1/4}{\rho^2} \right] f_0^{\text{in}}(\rho) = 0.
\end{equation}
For $\rho \to \infty$,
\begin{equation}
 \lim_{\rho \to \infty} f_0^{\text{in}}(\rho) \sim \sqrt{\rho} \sin(s_0 \log(\rho) + \phi_0).
\end{equation}
We determine the phase $\phi_0$ numerically, by choosing a boundary condition
\begin{equation}
 \left(\frac{f(\rho)}{\sqrt{\rho}}\right)' = \frac{\mu(\rho)}{\rho} \frac{f(\rho)}{\sqrt{\rho}},
\end{equation}
for small enough values of $\rho$, and by integrating the equation. The boundary condition ensures that we pick up the right solution. The result is $\phi_0 = s_0 \log(1.0918) + \alpha_0$. In order to find what value of $\kappa_*$ ($= \kappa$ in this limit) is induced by $r_s$, we have to match the inner and the outer solutions in the regime where $-r_s \ll R \ll a$. This yields that $-\kappa_* r_s = 1.0918$.  

Remarkably, this numerical value for $-\kappa_* r_s$ differs from the values obtained in Ref.~\cite{Petrov04} and \cite{Gogolin08}, which are $- \kappa_* r_s \approx 5$ and $- \kappa_* r_s \approx 5.3062$ respectively. It is unclear where the difference comes from, however, both papers use models different from the hyperspherical formalism that we used here, where we base our treatment of the effective range parameter on the descriptions in Refs.~\cite{Fedorov01} and \cite{Platter09}. More research is needed to resolve this discrepancy.

Now we are in a good position to compare our model with earlier work on the Efimov effect in a harmonic potential. In the work of Th{\o}gersen, Fedorov, and Jensen, the influence of the effective range is studied on the universal Efimov effect \cite{Thogersen08}. They considered a system with a Gaussian interaction potential, given by
\begin{equation}
V(r) = V_0 \exp(-r^2/R_0^2),
\end{equation}
where $V_0$ and $R_0$ are constants. The range $R_0$ has been chosen to equal $R_0=a_{\rm ho}/3965$.

\begin{figure}
\begin{center}
\includegraphics[width=7.5cm]{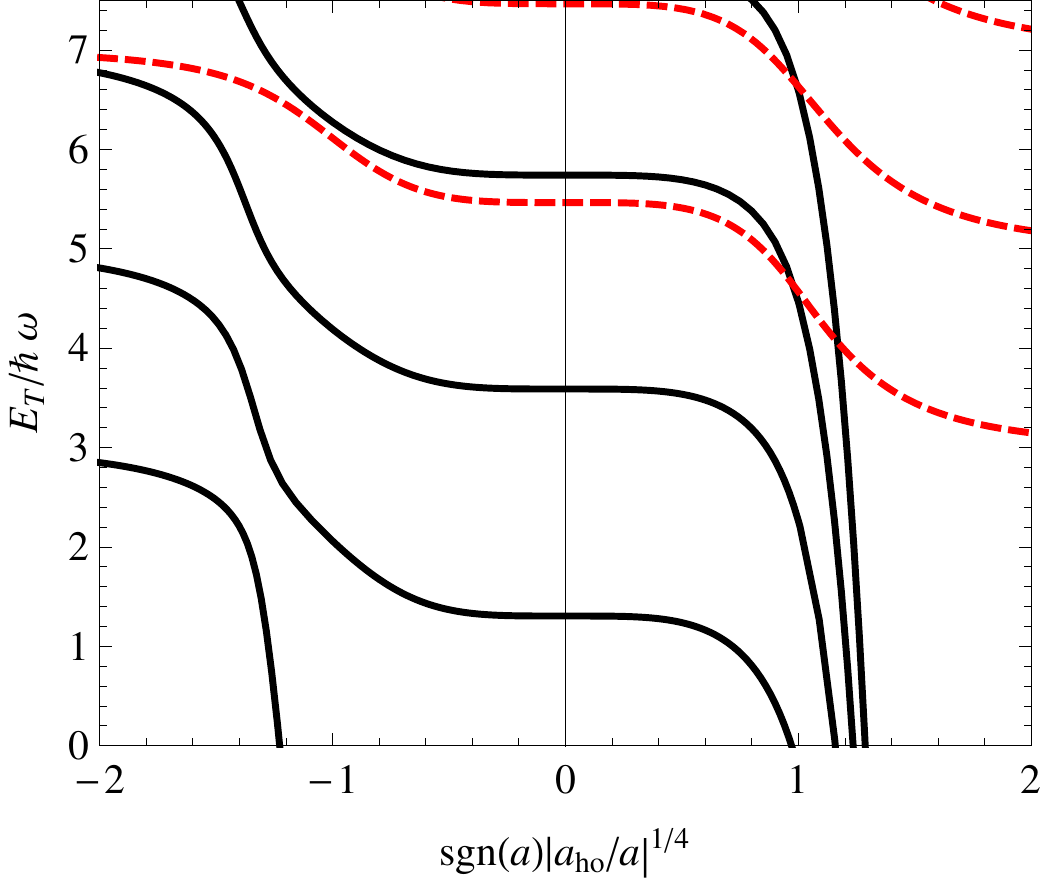}
\caption{Trimer solutions in a harmonic oscillator, for potential parameters corresponding to the paper of Th\o{}gersen {\it et al.}~\cite{Thogersen08}. Shown are the solutions above the three-particle threshold. The zeroth (first) channel solutions are given by black-solid (red-dashed) lines. The graph can be directly compared to Fig.~1 in this paper.}
\label{fi:topth}
\end{center}
\end{figure}

\begin{figure}
\begin{center}
\includegraphics[width=7.5cm]{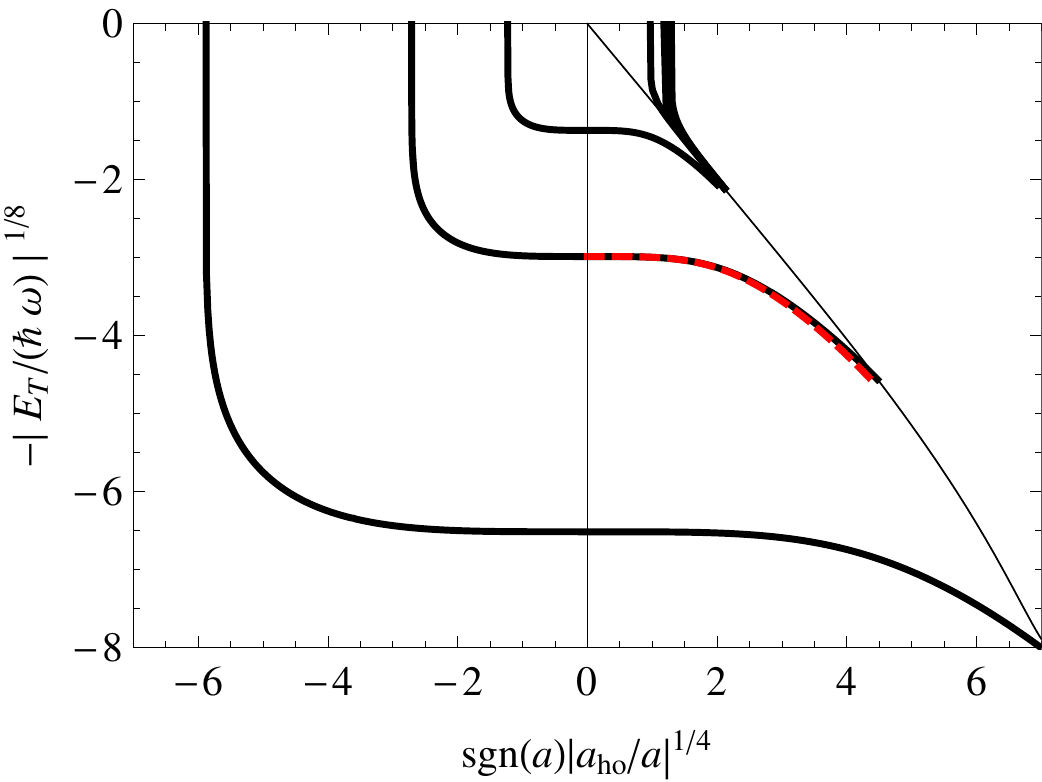}
\caption{Trimer solutions in a harmonic oscillator, for potential parameters corresponding to the paper of Th\o{}gersen {\it et al.}~\cite{Thogersen08}. Shown are the solutions below the three-particle threshold. The red line indicates a solution when the effective range is taken into account. Also indicated is the dimer-atom threshold, which includes effective range corrections. The graph can be directly compared to Fig.~1 in this paper.}
\label{fi:botth}
\end{center}
\end{figure}

Although the main purpose of Ref.~\cite{Thogersen08} is to show the influence of the effective range, their model contains a harmonic trap as well. In order to make a proper comparison with our model, we take $\kappa_* a_\text{ho} = 3.7$. The results are shown in Figs.~\ref{fi:topth} and Fig.~\ref{fi:botth}, and can be directly compared with Fig.~1 in Ref.~\cite{Thogersen08}. We observe that the numerical results are reproduced quite closely and we can assign the numerically found bound states to the different hyperspherical channels. We also observed that if we include the above corrections that come from the effective range, the comparison is better than results obtained without these corrections.

\section{Conclusion}

We studied the Efimov effect in a harmonic potential by using the hyperspherical formulation. In the first part of this paper, we studied the universal regime by only taking a large scattering length into account, and by making the adiabatic hyperspherical approximation in that we ignored the coupling between different hyperspherical channels. The model thus obtained is simple in the sense that the binding energies of the trimers result from an ordinary differential equation, with an appropriate boundary condition for the lowest channel encoded by the three-body parameter $\kappa_*$. Considering the lowest channel, we formulate an equation for a phase variable, the exponent of which is the radius in a spherical plot describing a surface of values in $(a^{-1},K,a^{-1}_{\rm
ho})$-space (inverse scattering length, wavenumber and inverse oscillator length) for which bound states exists. Other such surfaces are obtained by multiplying the distance to the origin by a factor $e^{\pi/s_0}$, and there is an accumulation point of surfaces near the origin, analogous to the Efimov effect in free space. In the limit $a_{\rm ho} \to \infty$, the Efimov effect in free space is recovered.

Interesting differences with the situation of three particles in free space can be observed when we focus at a fixed value of $a_{\rm ho}$. For parameter combinations where in free-space bound states would exist, the situation is very similar, except that the accumulation of energies only happens up to a typical energy of $\hbar \omega$. The rest of the Efimov states are "pushed out" into the trimer continuum (which is of course only a continuum in free space). Higher hyperspherical channels give rise to additional bound states, and by taking those into account as well we obtain results very close to numerical simulations of three-particle systems~\cite{Thogersen08}.  Better qualitative results are obtained by approximating the coupling between the hyperspherical channels, which lifts the degeneracies and creates avoided crossings. The lifetime of the trimers is highest for states with a continuum character, i.e.~in the region which we would denote as trimer continuum in free space. The reason for this is the small
overlap of these states with deeply bound states. Experimental study of Efimov-like states, however, could be feasible close to the transition from this continuum region to the trimer region, where the linewidth is still quite small.

For a more general Efimov model that allows for deviations from the universal regime, a natural step is to improve on the approximation of the hyperangular problem by taking into account the effective range. By doing this, we were able to find a relation between $\kappa_*$ and the effective range $r_s < 0$ if $a \gg -r_s \gg R_0$, where $R_0$ is a typical range of the potential. We obtain the same qualitative behavior as Ref.~\cite{Petrov04} and \cite{Gogolin08} where different descriptions were used, however, the actual value of $\kappa r_s$ is also very different. It appears that the actual value for this relationship depends strongly on the choice of the model and the corresponding approximations to describe the three-particle physics, and more research is needed to understand this discrepancy. In a different parameter regime for $a$, $r_s$ and $R_0$, such a relationship between $\kappa_*$ and $r_s$ is not necessary. Then we have in addition to $r_s$, the three-body parameter $\kappa_*$ in the boundary condition for the hyperradial problem to our disposal. By comparing to numerical calculations we find that this description indeed improves the predicted trimer energies.


\end{document}